# Development of Systematic Image Preprocessing of LAPAN-A3/IPB Multispectral Images

Patria Rachman Hakim, A Hadi Syafrudin, Sartika Salaswati, Satriya Utama, Wahyudi Hasbi
Satellite Technology Center
National Institute of Aeronautics and Space (LAPAN)
Bogor, Indonesia

*Abstract*—As of any other satellite images, LAPAN-A3/IPB multispectral images suffered from both geometric and radiometric distortions which need to be corrected. LAPAN as satellite owner has developed image preprocessing algorithm to process raw image into systematically corrected image. This research aims to evaluate the performance of the developed algorithm, particularly the performance of lens vignetting and band co-registration correction as well as the performance of direct image georeferencing. Lens vignetting distortion occurs on image was corrected by using pre-flight calibration data, while calculation of direct georeferencing was done by using satellite metadata of satellite position and attitude. Meanwhile, band co-registration correction was conducted based entirely on the image being processed using image matching approach. Based on several results and analysis which have been done, lens vignetting effects on image can be suppressed significantly from about 40 percent down to 10 percent, band coregistration error can be reduced to below 2-3 pixels in average, and the calculated direct georeferencing has 3000 meter accuracy. The results show that the developed image preprocessing algorithm has moderately good performance to process LAPAN-A3/IPB multispectral images.

## I. INTRODUCTION

In order to develop sovereign national capability in aerospace technology, National Institute of Aeronautics and Space (LAPAN) has developed and launched three satellites since 2007. LAPAN-A1/TUBSAT was the first satellite launched in 2007, which has sun synchronous orbit and analog video camera as its main payload [1]. The satellite still exists in orbit and can be well controlled, but with the age of about 11 years, the payload has lost its capability. Its successors, LAPAN-A2/ORARI and LAPAN-A3/IPB were launched in 2015 and 2016, respectively. LAPAN-A2 is equatorial microsatellite with earth surveillance mission, which has 4 meter resolution digital camera and Automatic Identification System (AIS) as main payloads, as well as Voice Repeater (VR) and Automatic Packet Reporting System (APRS) as supporting payloads [2]. In the other hand, LAPAN-A3 is polar sun synchronous microsatellite with remote sensing mission, which has multispectral pushbroom imager and AIS as the main payloads, supported by several scientific experimental payloads such as a 3 meter resolution digital camera, magnetometer, and thermal imager [3]. Currently, both satellites are operating nominally and have been producing images and AIS data regularly which are used by numerous institution in Indonesia [4][5][6].

Multispectral imager of LAPAN-A3 satellite has four color channels of blue, green, red and near-infrared (NIR) with 15 meter resolution, 120 km swath-width and 21 days revisit time [7]. The imager has been producing Indonesia coverage images for 300 thousand kilometer square daily. However, as of any other satellite remote sensing images, images produced by LAPAN-A3 multispectral pushbroom imager suffered from several geometric and radiometric distortions. One of most notable radiometric distortions occurs on satellite image is vignetting effect caused by imperfect lens geometry which causes center of image looks brighter than the edge parts [8][9]. In the other hand, one of most notable geometric distortions is band co-registration differences between all of image bands, which is caused by different position and orientation of each band detector relative to the lens center [10][11]. These two distortions significantly degrade the quality of raw image captured, therefore should be corrected in order to produce high quality of end product image, both in terms of geometry and radiometry point of view. Furthermore, since the images might not always come from nadir pointing observation, there is uncertainty corresponds to geospatial location of image captured, which will affect the performance of further image post-processing for numerous remote sensing applications. Therefore, image georeferencing is needed to determine image geolocation, which can be done manually by using ground control points or systematically by using satellite sensor metadata [12][13].

To guarantee that LAPAN-A3 multispectral imager could produce high quality imagery, LAPAN as satellite owner has developed systematic image preprocessing algorithm, which is implemented on a dedicated software. In general, the developed software processes raw image captured into systematically corrected image, by sequentially performing several stages of image data processing. Several previous researches have been done to develop the algorithms, but all of them were conducted either individually or not using the actual LAPAN-A3 images. Characterization of vignetting distortion on LAPAN-A3 multispectral images had been conducted previously, however the work was based only on pre-flight calibration data, not the actual satellite images captured [14]. Works related band co-registration correction perhaps have been conducted more thoroughly, either by using combined of edge detection and image matching approach [15] or by using attitude modeling approach [16]. However, the performance was not evaluated quantitatively for general LAPAN-A3 images. Meanwhile, research work about systematic direct image georeferencing of LAPAN-A3 multispectral images had only be conducted in simulation before the satellite was launched [17].

This research aims to develop comprehensive systematic image preprocessing of LAPAN-A3/IPB multispectral image and to analyze the overall performances of the developed algorithm with respect to three previous above mentioned aspects, which are lens vignetting and band co-registration as well as direct image georeferencing. The performance will be evaluated quantitatively based on numerous of the actual LAPAN-A3 multispectral images captured. In this research, performance metrics used to evaluate the algorithm are brightness ratio between image edge region to image center for vignetting distortion and corresponding pixel position differences between each of image bands for band co-registration distortion. Direct image georeferencing has more straightforward performance metric, which is difference between the calculated image geospatial location to the actual image location from manually georeferenced images by using proven map. The overall processing time needed by the software to execute the whole algorithm is also evaluated to provide an additional performance measurement. Several possible future researches in order to further improve the quality of LAPAN-A3 multispectral image preprocessing is also explored, to guarantee that LAPAN satellites always deliver the best quality images to the end-user.

The paper is organized as follows. Section II will give description of the developed algorithm with some theoretical backgrounds of satellite image preprocessing, while section III will discuss about the developed algorithm performances on LAPAN-A3 multispectral satellite images. Finally section IV will give the conclusion of this research.

## II. METHODOLOGY

In general, this research consists of three important stages. First stage was reviewing previous systematic preprocessing algorithm of LAPAN-A3 multispectral image, followed by improving the algorithm in several aspects. Second stage was an assessment of each submodul algorithm performance as well as the overall performance of the developed algorithm. Three submodul of preprocessing algorithms were heavily analyzed, i.e.: lens vignetting and band co-registration correction as well as direct image georeferencing calculation. Finally, third stage of this research explored about potential aspects which can be enhanced in order to further improve the quality of systematic image preprocessing software for LAPAN-A3 multispectral images, either in terms of algorithm improvement or practical implementation of the software. Figure 1 shows general flowchart for methodology which is used in this research.

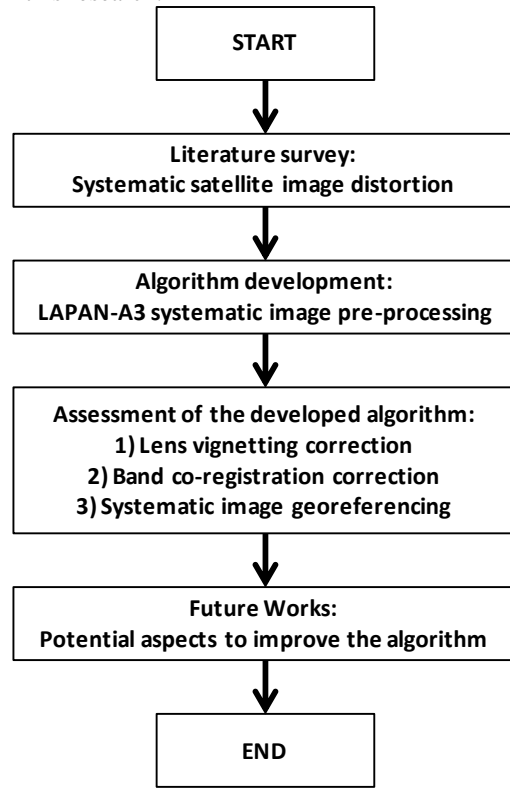

Figure 1. Methodology used in this research

Vignetting distortion occurred on satellite images mainly caused by the imperfect of lens geometry so that incoming light does not refracted uniformly across all direction as well as by the different incoming angle with respect to each detector. The distortion causes the image center to look brighter than image corner, or both sides of the image in the case of pushbroom imager. This might cause two identical objects located on different side of the image might have different digital number value, which in turn would cause significant performance degradation of image post-processing applications. Lens vignetting distortion on the images generally can be corrected by using calibration data which were obtained before the satellite was launched, where the calibration data used usually consist of measurement of uniform light source by using integrating sphere, measurement of detector dark current under perfect dark environment, and measurement of imager spectral response by using monochromator [18]. The corrected image can be then obtained by applying these calibration data into uncorrected image based on imager radiometric model. However due to significant imager temperature different between earth and space, dark current calibration data obtained before the launch could not be used properly to correct the vignetting distortion. Therefore in this research, dark current calibration data was replaced by dark image obtained from deep space or night ocean observation. Figure 2 shows pre-flight calibration data of LAPAN-A3 multispectral imager, showing measurement of imager spectral response as well as flat-field measurement of integrating sphere.

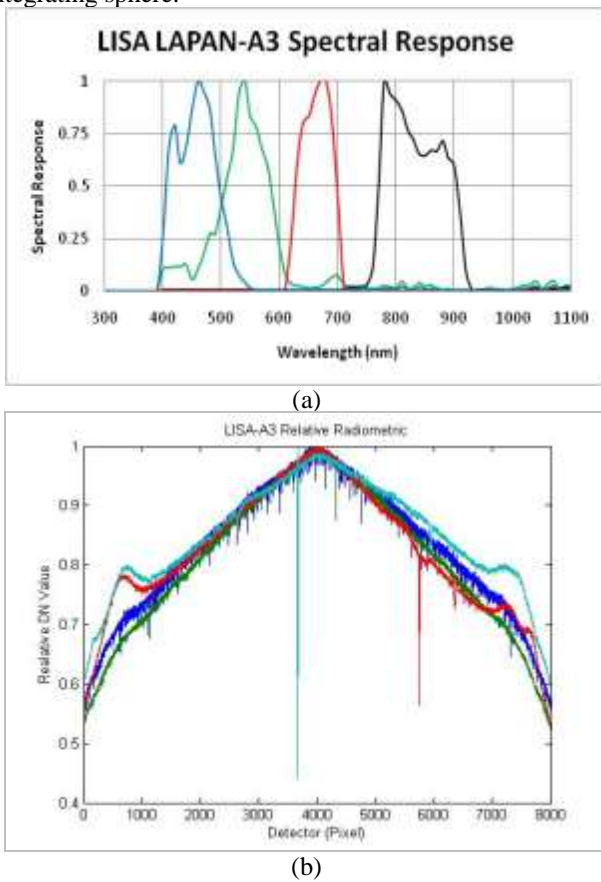

(a)

(b)

Figure 2.  Pre-flight radiometric calibration data of LAPAN-A3 multispectral imager; (a) detector relative spectral response, (b) detector relative vignetting response

Since the multispectral imager is pushbroom type of imager, then vignetting correction algorithm is applied for each image line. For given relative vignetting calibration data of $R_i$ and dark image observation data of $D_i$, where $i$ represents pixel location on detector, then the corrected image pixels of $Y_i$ can be obtained from distorted raw image pixels of $X_i$ by using following simple matrix equation:

$$Y_i = R_i * ( X_i - D_i ) \qquad (1)$$

Meanwhile, band co-registration distortion occurred on satellite images directly caused by position and orientation difference between each of band detectors with respect to lens center, which causes pixels in the same position on each detector might capture different objects. The distortion causes objects in one image band do not aligned with the same objects on other band images, which is visually similar to blurring effect on the image. Significant co-registration distortion on the image might produce severe information interpretation error, for example Normalized Difference Vegetation Index (NDVI) assumes that the digital number value of the corresponding red and NIR band come from the same object. Band co-registration distortion occurred on the image is usually corrected by either manual approach which is laborious and also time consuming [5] or by using rigorous imager geometry model approach which needs some accurate imager geometric parameter [19]. Therefore to overcome these constraints, band co-registration distortion on LAPAN-A3 multispectral image is corrected by using image matching approach. The correction method utilizes combination of edge detection and intensity-based image matching algorithm, followed by distortion modeling and image resampling [15]. The concept is based on the use of canny edge detection to perform high-pass filtering [20][21] and fast-fourier transform to perform two-dimensional image cross-correlation approach of image matching [22][23]. In general, the algorithm works very well but sometimes it fails to produce accurate results due to the existence of outlier data on distortion modeling process. In this research, more robust outlier removal algorithm is used by using satellite attitude information. Figure 3 shows existing algorithm of band co-registration correction currently used to process LAPAN-A3 multispectral image [15].

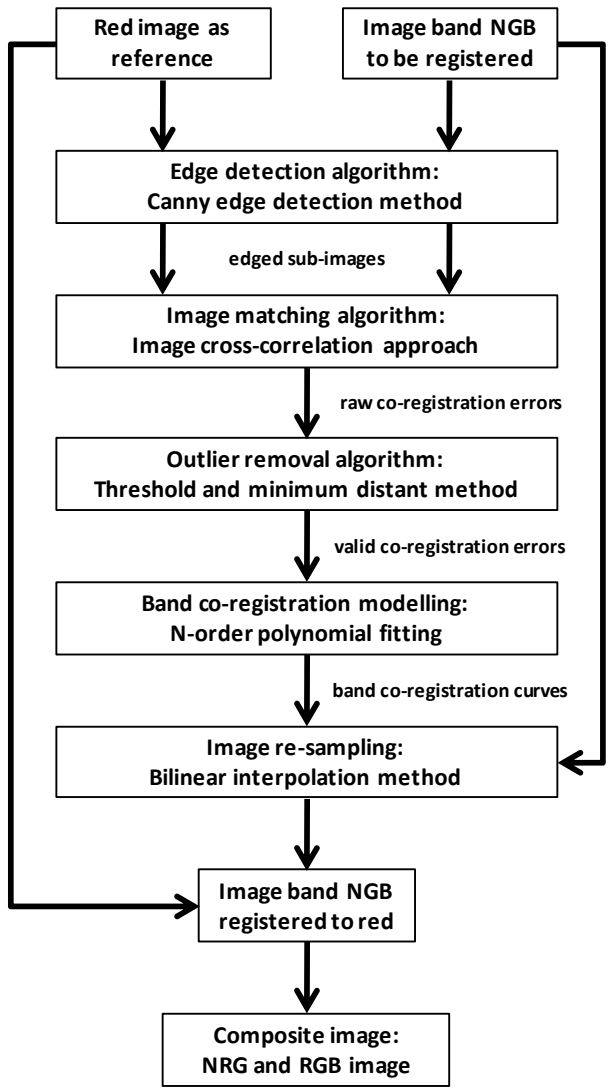

Figure 3. Algorithm of band co-registration correction of LAPAN-A3 multispectral images [15]

Having two most major distortions been corrected, the third step in this research was calculating location of the image on the earth surface by using satellite metadata of position and attitude while imaging, which is commonly referred as image direct georeferencing. By knowing both satellite position and attitude while the image was taken, the projection of the image in earth surface can be calculated [24]. Image georeferencing can also be done without satellite information by using ground control point (GCP) approach, which could produce more accurate result, but it also needs significant human interaction and also time consuming [12]. In general, existing LAPAN-A3 multispectral image direct georeferencing can be divided into two parts, where first part was calculating satellite position and attitude from satellite system time and star tracker sensor (STS) raw data by using SGP4 orbit model and celestial coordinate transformation relationship [25][26]. In second part, after determining satellite position and imager attitude while imaging, image pixels position on earth surface can be calculated by using pinhole camera model and collinear projection relationship utilizing orbit parameter such as satellite altitude as well as imager parameters such as focal length and detector width [17]. Figure 4 shows general flowchart of LAPAN-A3 systematic image direct georeferencing.

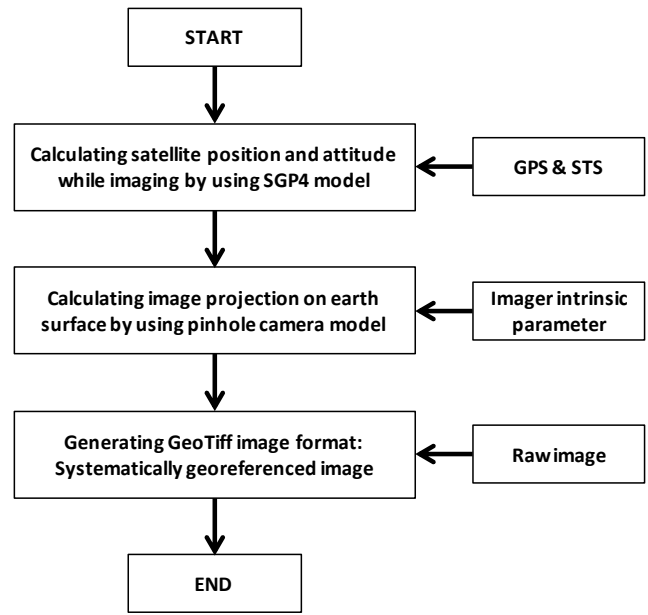

Figure 4. Algorithm of image direct georeferencing of LAPAN-A3 multispectral images

### III. RESULTS AND DISCUSSION

The developed algorithm has been validated on more than 1000 actual LAPAN-A3 multispectral images, which have been taken since its launch in 2016. Based on these processed images, each sub-algorithm will be analyzed next, which are radiometric vignetting correction, geometric band co-registration correction and also systematic image direct georeferencing.

#### A. Radiometric Vignetting Correction

Basically, vignetting correction algorithm will transform uncorrected image into corrected image with more uniform color image across horizontal direction when the imager observe relatively similar objects. Therefore, to analyze the performance of the algorithm, images with highly uniform objects such as desert, savana field, karst field, and ocean are used. Figure 5 shows radiometric vignetting correction result for LAPAN-A3 multispectral image of desert and coastal on northern Egypt, while figure 6 shows correction result for image on southern Pakistan.

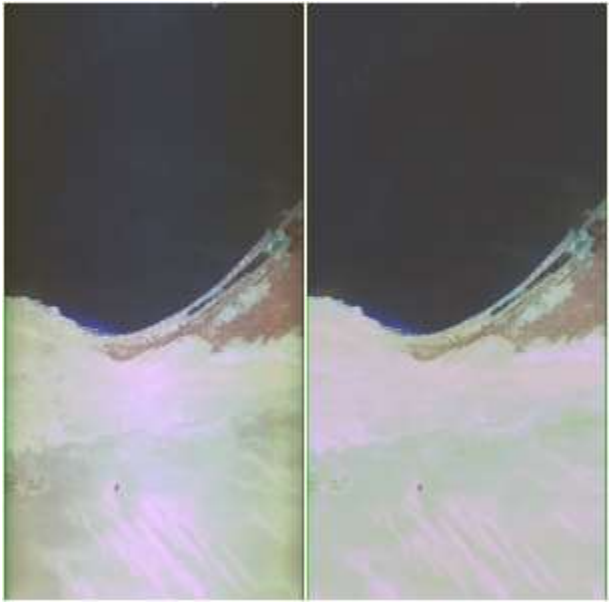

Figure 5. Vignetting correction on image of Port Alexandria, Egypt, taken on 13 October 2016; uncorrected (left) and corrected (right)

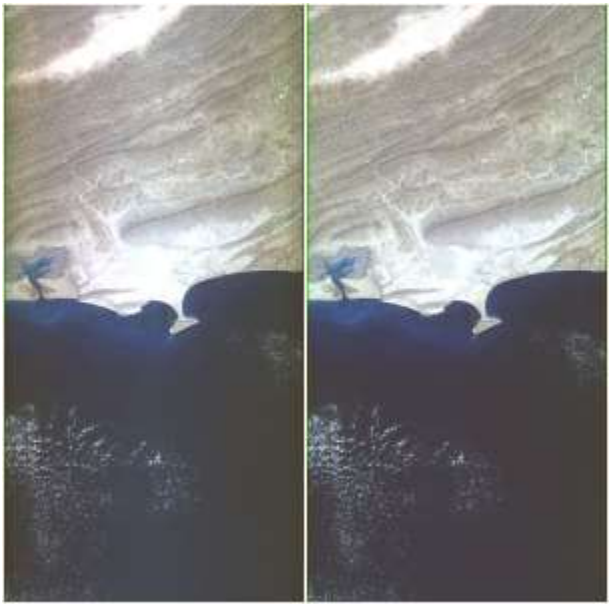

Figure 6. Vignetting correction on image of southern Pakistan, taken on 11 March 2018; uncorrected (left) and corrected (right)

From both figure 5 and figure 6, it can be seen that the algorithm successfully minimize the vignetting effect on the both images. It can be clearly seen visually that desert area of northern Egypt on corrected image has significant uniformity improvement compare to desert on the uncorrected image. The sea and coastal area on both Egypt and Pakistan images are also visually well corrected, where the uniformity of both corrected image are much better. To further quantitize the quality performance of the algorithm, figure 7 shows average relative digital number (DN) value for each image line for both uncorrected and corrected of Egypt image, for both low and high radiance area. It can be seen that vignetting effect on uncorrected image can be reduced from around 40 percent down to about 10 percent. Note that the curves on the figure are second order polynomial approximation of actual relative digital number curves, in order to simplify the analysis of algorithm performance.

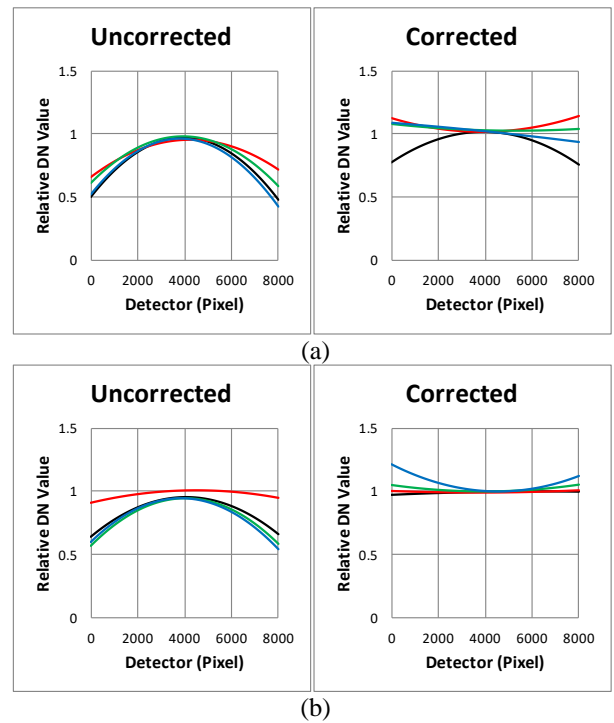

Figure 7. Performance of vignetting correction algorithm on Egypt image, which is applied on; (a) low radiance area, (b) high radiance area

The performance of the vignetting correction algorithm has also been evaluated based on the actual relative digital number by using area uniformity approach, which can be estimated from standard deviation value. An image area is said to be highly uniform when majority of its digital number values has more or less similar value, which means that it will have relatively low standard deviation value. Based on previous Egypt image, it is found that average of standard deviation of several selected lines on each image band before correction is about 15 percent, and then reduced to about 10 percent after vignetting correction is applied, which shows that corrected image has slightly higher uniformity compare to the uncorrected image.

*B. Geometric Band Co-registration Correction*

The developed band co-registration correction algorithm will transform NIR, green and blue band images so that any object on all four image bands are aligned to each other. To analyze the performance of band co-registration correction algorithm, residual band co-registration must be determined, which needs detail object to be analyzed. Therefore, images with highly structured geometric objects are used, such as urban area where consists of numerous buildings and any other man-made objects. Figure 8 shows correction result of geometric band co-registration for LAPAN-A3 multispectral image of urban area on Indianapolis of United States, while figure 9 shows correction result for image of urban area on Kuwait City.

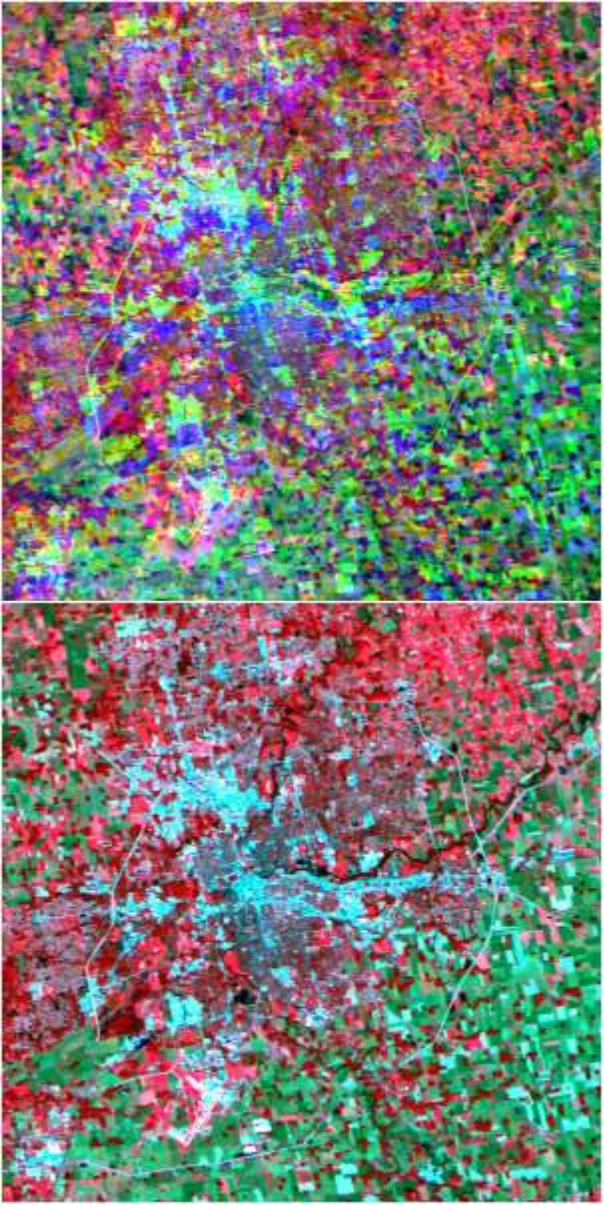

Figure 8.  Band co-registration correction on image of Indiana, US, taken on 27 September 2016; uncorrected (top) and corrected (bottom)

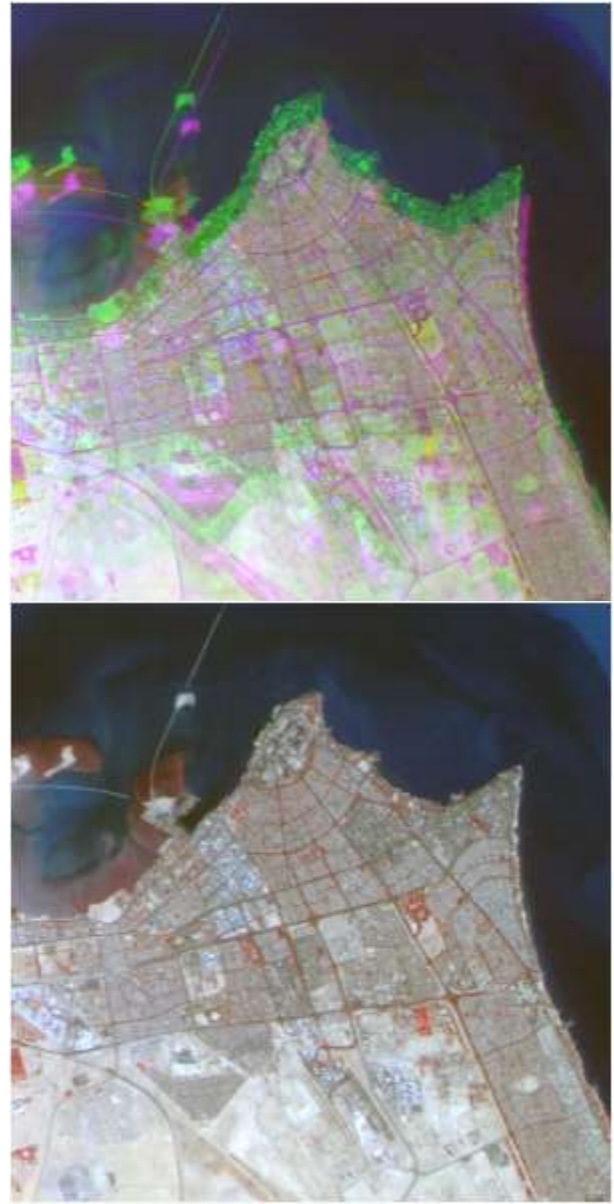

Figure 9.  Band co-registration correction on image of Kuwait City, taken on 7 March 2018; uncorrected (top) and corrected (bottom)

Visually, it can be clearly seen that geometric quality of both images has increased significantly due to geometric band co-registration correction applied. From both figure 8 and figure 9, blurring-like effect on uncorrected images can be minimized, where many objects such as buildings, roads and bridges can be identified easier, since on the corrected images, these objects have already aligned to each other on each image band. Quantitative performance of the algorithm has also been analyzed, by manually determining residual band co-registration error of the corrected images, by using around 50 control points for each band registration. Based on several images analyzed, it is found that average accuracy of the developed band co-registration correction algorithm is about 2-3 pixels error.

It is also found that band co-registration distortion occurs on the image is highly influenced by satellite attitude while imaging [16]. Images captured while satellite attitude was on nadir oriented will have generally lower band co-registration distortion compare to the images taken on off-nadir oriented satellite attitude. Figure 10 shows images of Bali island taken from both nadir and off-nadir satellite position, while figure 11 shows comparison of the resulted band co-registration error of each image band with respect to red band image. It can be seen that images taken from off-nadir viewing angle have more significant band co-registration distortion. Figure 11 also shows that most of band co-registration distortion has non-linear characteristics, both in vertical and horizontal direction.

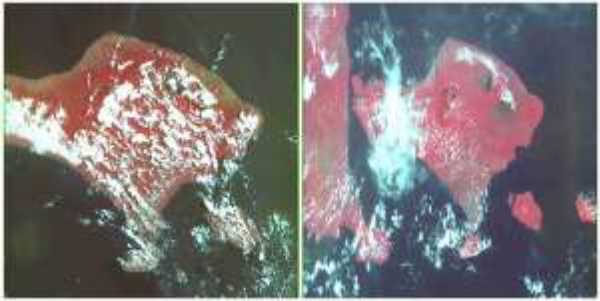

Figure 10. Observation of Bali island; captured from nadir satellite position (left) and captured from off-nadir satellite position (right)

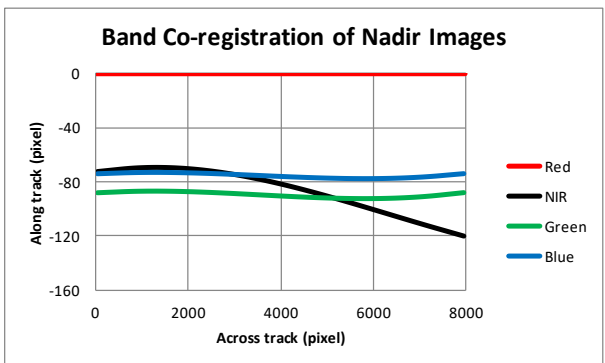

(a)

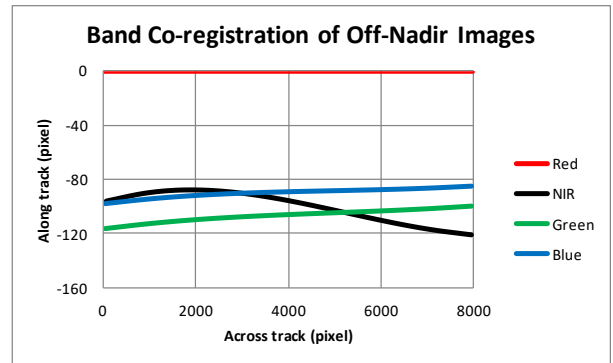

(b)

Figure 11. Example of band co-registration distortion of LAPAN-A3 multispectral images; (a) captured from nadir satellite position, (b) captured from off-nadir satellite position

## C. Systematic Image Georeferencing

As already mentioned earlier, systematic georeferencing of LAPAN-A3 multispectral images is conducted by using satellite metadata of satellite system time from GPS and also satellite attitude form star tracker sensor. Figure 12 and 13 show the resulted systematic image georeferencing on two LAPAN-A3 multispectral images of Central Java and Bali island, respectively.

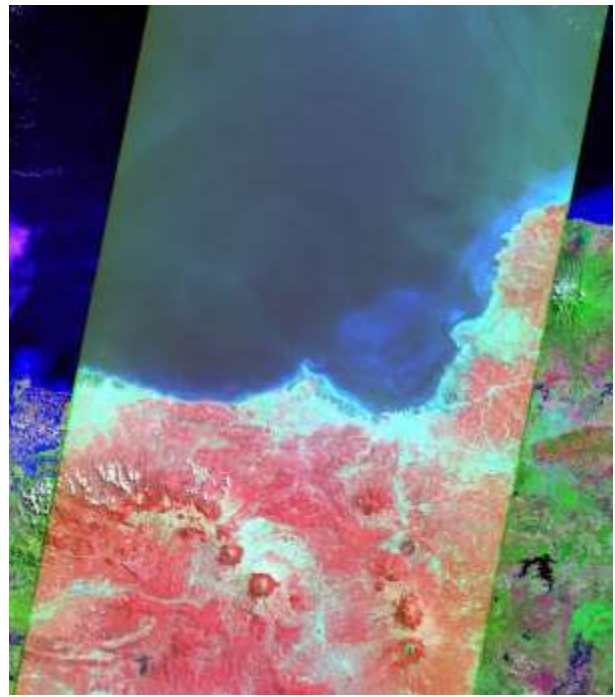

Figure 12. Direct georeferencing on image of Central Java, Indonesia, taken on 5 May 2018

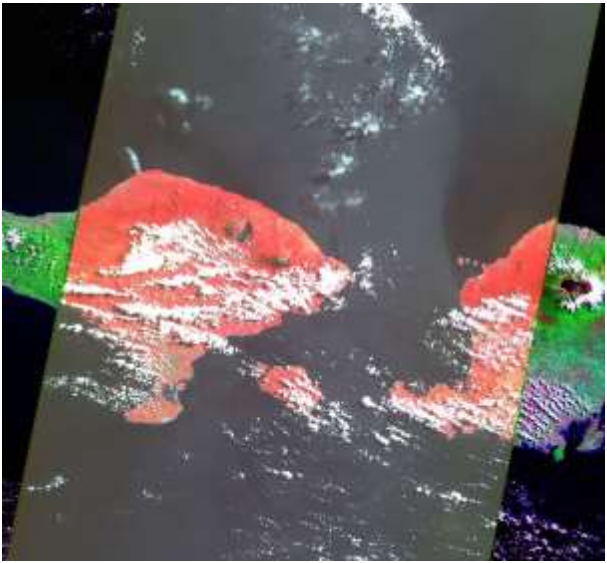

Figure 13. Direct georeferencing on image of Bali and Lombok islands, Indonesia, taken on 4 July 2018

It can be seen that the developed algorithm successfully calculates geolocation of the images, where image on figure 12 and 13 has 8.18 km and 7.85 km accuracy, respectively. By further analyzing several other images, it is found that average accuracy of the image georeferencing is about 11.34 km, which can be divided into 3.80 km accuracy on across-track image direction and 10.69 km accuracy on along-track image direction. Based on these images, figure 14 shows statistic of the across-track error produced, which has normal distribution with mean value of 3.57 km error. By assuming that normal distribution has zero mean value, this 3.57 km mean value can be considered as an offset error which is produced systematically by the satellite system. In this case, misalignment of the imager and star tracker sensor as attitude sensor can be considered as the main source, particularly in roll imaging axis for across-track error. After recalculating image geolocation with 0.401 roll angle offset, corresponds to 3.57 km earth distance, average of across-track accuracy of image georeferencing can be improved to 2.03 km.

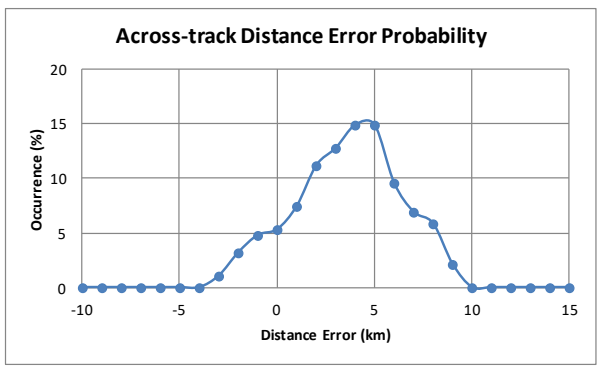

Figure 14. Statistical georeferencing error in across-track direction

In the other hand, significant image georeferencing error in along-track direction is caused by imager misalignment to star tracker sensor and satellite system time error. Figure 15 shows influence of the system time error to along-track error produced. Recalculating image geolocation with 1.09 pitch angle offset and system time error information, average of along-track accuracy of the image georeferencing can be improved to 2.20 km. Combination of both across-track and along-track accuracy improvement produces overall image georeferencing accuracy of 2.99 km.

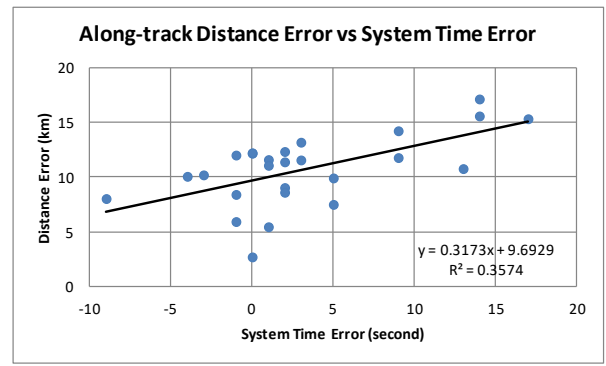

Figure 15. Systematic georeferencing error in along-track direction

*D. Practical Implementation*

From several previous analyzes, it can be said that the developed algorithms has moderate performance, where lens vignetting algorithm could reduce vignetting effect from 40 down to 10 percent, band co-registration algorithm has about 2-3 residual pixel error and systematic image georeferencing algorithm has about 3000 km accuracy. To further assess the overall performance of systematic image pre-processing of LAPAN-A3 images, figure 16 shows processing time needed to process a single image, based on the image size. It can be seen that average processing time needed is about 10 minutes for each image processed, which can be considered fast to process an uncorrected image into a systematically corrected and georeferenced image.

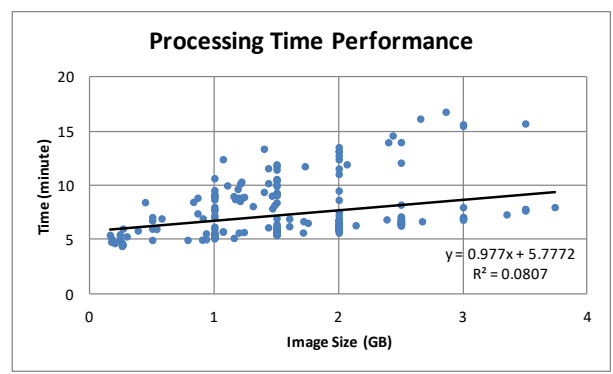

Figure 16. Processing time needed to process a raw image

In general, for most of the processed raw images, half of processing time needed was spent to execute image band co-registration correction algorithm, especially image matching algorithm. Although the processing time of overall algorithm can be considered fast enough, an improvement is always welcome. This can be accomplished by reducing number of matching points that are used in image matching algorithm. However, the performance of band co-registration algorithm might be degraded, therefore further investigation regarding this matter should be conducted in order to improve overall performance. The choice of polynomial models of each pair of band co-registration and further improvement of outlier removal method should also be conducted next to improve the accuracy of band co-registration algorithm.

Meanwhile, in order to improve the performance of systematic image georeferencing algorithm, determination of imager misalignment to star tracker sensor should be done more thoroughly. Satellite system time error characterization should also be determined precisely since along-track error of the resulted systematic image georeferencing, in general, has the biggest part in determining overall accuracy. With inclusion of both imager misalignment and satellite system time in the algorithm, accuracy of about 1000 meter could be expected. This residual accuracy error is mostly caused by satellite nutation effect in orbit, which causes LAPAN-A3 satellite to have sinusoidal oscillation in roll imaging axis with around 0.28 degree amplitude and 73 seconds period characteristic [27].

## IV. Conclusion

In order to produce high quality image, LAPAN has developed image pre-processing software to process raw LAPAN-A3/IPB multispectral images into systematically corrected images. The correction algorithm mainly consists of lens vignetting correction, band co-registration correction, and direct image georeferencing calculation. The software has been validated by using more than 1000 images which have been captured since satellite launch on June 2016. Lens vignetting distortion on images, which was corrected using pre-flight calibration data, can be reduced from about 40 percent down to 10 percent. Band co-registration distortion, which was corrected by using image matching algorithm, can be minimized for residual 2-3 pixel error. Meanwhile, image georeferencing by using satellite position and attitude data has around 3000 meter accuracy. Processing time needed to process an uncorrected image can be considered very fast, about 10 minutes in average, allowing fast distribution to the end user.

## Acknowledgment

The authors would like to thank Mr. Mujtahid as Director of Satellite Technology Center of LAPAN as well as Mr. Suhermanto as former director for their support so that this research can be well completed. The authors would also like to thank Ministry of Research, Technology and Higher Education of the Republic of Indonesia for their support with INSINAS 2018 funding.